\documentclass{amsart}
\newenvironment{nouppercase}{%
  \renewcommand{\uppercasenonmath}[1]{}}{}
\usepackage{enumerate,mathrsfs}
\usepackage{amssymb}
\usepackage{amsmath}
\usepackage{amsfonts}
\usepackage{braket}
\usepackage{color}
\usepackage{graphicx}
\usepackage{tikz,pgffor}
\usepackage{rotating}
\usetikzlibrary{calc, shapes, matrix,snakes,positioning,backgrounds,arrows,automata,backgrounds,fit,decorations.pathreplacing}
\usetikzlibrary{shadows}
\usepackage{tkz-graph}
\renewcommand{\subset}{\subseteq}

\DeclareMathOperator{\Aut}{Aut}

\DeclareMathOperator{\rank}{rank}
\DeclareMathOperator{\spn}{span}

\theoremstyle{definition}
\newtheorem{defn}{Definition}
\newtheorem{example}{Example}
\theoremstyle{theorem}
\newtheorem{theorem}{Theorem}

\title{\text{Quantum Probabilistic Spaces on Graphs for Topological Evolutions }}
\author{Radhakrishnan Balu}
\address{Computer and Information Sciences Directorate,\\ Army Research Laboratory, 
Adelphi, MD, 21005-5069, USA.}
\email{radhakrishnan.balu.civ@mail.mil}
\address{Department of Mathematics, University of Maryland, College Park, MD 20742}
\email{rbalu@math.umd.edu}
\begin{document}
\begin{nouppercase}
\maketitle
\end{nouppercase}
\begin{abstract}
We start with the consideration of fusion rules of anyonic particles evolving on a 2D surface and the a hypergroup comes with it to construct entangled quantum Markov chains. The fusion rules induce an association scheme with  Krein parameters and their duals the intersection numbers. One useful way to think of the schemes as regular  graphs encoding the paths of possible quantum walks (automorphisms). We consider braid $B_3$ that describes the unitary dynamics of the anyons as the automrphism subgroup of the graphs. The dynamics induced by the fusions (and the adjoint splitting operations) may be viewed as the chain evolving on a growing graph and the braiding as automorphisms on a fixed graph. In our quantum probability framework infinite iterations of the unitaries, which can encode algorithmic content for quantum simulations, can describe asymptotics elegantly if the particles are allowed to evolve coherently for a longer period. We will define quantum states on the Bose-Mesner algebra which is also a von Neumann algebra as well as a Frobenius algebra to build the quantum Markov chains providing another perspective to topological computation. 

\end{abstract}
\section {Introduction}
 2D anyonic computations can be described using unimodular tensor categories (UMC),  which is equivalent to a 2 + 1 topological quantum theory (TQFT) \cite {Turaev1994}, or by the cobordism hypothesis \cite {Freed2013}. In an earlier work \cite {Radbalu2020} we studied association schemes (key notions and typical examples are summarized in appendix) that may be thought of as adjacency matrices of graphs to derive mathematical structures relevant to quantum processes. We considered finite groups that induce association schemes with parameters Krein and intersection numbers that are duals. The adjacency matrices $A^n$ give rise to a von Neumann algebra (Bose-Mesner) that can encode paths of quantum walks using intersection numbers and in the dual picture the Krein parameters can represent particle collisions which can encode fusion rules of anyons and the comparison between the frameworks is summarized in Table 1.  
 \tikzset{ 
    table/.style={
        matrix of nodes,
        row sep=-\pgflinewidth,
        column sep=-\pgflinewidth,
        nodes={
            rectangle,
            draw=green,
            align=center
        },
        minimum height=4.0em,
        text depth=0.5ex,
        text height=2ex,
        nodes in empty cells,
        every even row/.style={
            nodes={fill=gray!20}
        },
        column 1/.style={
            nodes={text width=14em,font=\bfseries}
        },
        row 1/.style={
            nodes={
                fill=blue!20,
                text=red,
                font=\bfseries
            }
        }
    }
} 
 \begin{figure*}
\centering
\begin{tikzpicture}

\matrix (first) [table,text width=14em]
{
UMC  & Anyonic system & Association Schemes \\
simple object & anyon & dualizable adjacency matrix  \\
label & anyon type or anyonic charge &  B-M algebra element\\
tensor product & fusion & Schur multiplication  \\
fusion rules & fusion rules & Krein fusion rules  \\
triangular space  $V_c^{ab}$ or $V_{ab}^c$ & fusion/splitting space & triangular space $V_c^{ab}$ or $V_{ab}^c$ \\
dual & antiparticle & matrix adjoint \\
birth/death & creation/annihilation & birth/death \\
mapping class group representations & generalized anyon statistics & generalized anyon statistics \\
nonzero vector in V (Y ) & ground state vector & unit element \\
unitary F-matrices & recoupling rules & unitary F-matrices\\
twist $\theta_x = e^{2\pi sx}$ & topological spin & q-deformed state \\
colored braided framed trivalent graphs & anyon trajectories & trivalent graphs\\
quantum invariants & topological amplitudes & probability amplitudes \\
};
\end{tikzpicture}
 \caption {Table 1: Comparision between UMC, Anyonic systems, and Association schemes. Extension of the table in Wang{'} work \cite {Wang2013}}
 \centering
 \end {figure*}
Another way to make a connection between UMC and association schemes is to start with a finite group. Then, consider the UMC and the fusion rules it induces, see \cite {Wang2013} for how to generate the parameters S, $\theta$ etc, and at the same time the association scheme the group leads to along with the Krein parameters. This roughly produces an equivalence between the formalisms. 

The organization of the manuscript is as follows: In section 2 we discuss the Ising anyons  in terms of association schemes. We then define quantum Markov chains for this system using our earlier work on B-M algebras based on the usual matrix product we denote by $\bullet$ and Hadamard operations (Schur multiplication) we indicate by the symbol $\circ$, and the tensor product between matrices with the usual symbol $\otimes$. We use the same symbol $\circ$ to indicate Schur multiplication when we extend this operation from matrix multiplication to the corresponding tensor operation. We then treat the example of Fibonacci anyons and build entangled versions of quantum Markov chains again based on our earlier work. We conclude the discussions summarizing the results and outlining the future work. To explore deeper connection between Markovian evolutions and topological quantum field theories we recommend the work of Levy \cite {Levy2011}

\begin {example}
Let us start with the fusion rules of the Ising model with the majorana fermion and the non-Abelian anyon $\psi, \sigma, 1$ as
\begin {align}
\sigma \times \sigma &= 1 + \psi. \\
\sigma \times \psi &= \sigma. \\
\psi \times \psi &= 1. 
\end {align}
There is a standard way to derive the F and R matrices for this anyonic systems:
\begin {align}
F^\sigma_{\sigma \sigma\sigma} &= \pm \frac {1}{\sqrt{2}} \begin{bmatrix} 1 & 1 \\ 1 & -1 \end {bmatrix}.  
\end {align}
The R entries are determined by F as 
\begin {align*}
R^1_{\sigma \sigma} &= \pm iR^{\psi}_{\sigma \sigma}. \\
R^{\psi}_{\sigma \sigma} &= \pm e^{-i\frac{3\pi}{8}}. \\
R^{\sigma}_{\sigma 1} &= 1. \\
R_{\sigma\sigma} &= e^{\frac{i\pi} {8}} \begin{bmatrix} 1 & 0 \\ 0 & i \end {bmatrix}.
\end {align*}
The Braiding matrix can be written in terms of R and F as $B = FR^2 F^{-1}$.
Now, the fusion rules can be expressed as  
\begin {align}  \label {FusionRules}
N^1_{\sigma \sigma} &= 1. \\  \nonumber
N^{\psi}_{\sigma \sigma} &= 1. \\ \nonumber
N^{\sigma}_{\sigma 1} &= 1. \\  \nonumber
N^{\sigma}_{\sigma \psi} &= 1. \\ \nonumber
\end {align}
The dimension of an anyon can be calculated from the fusion rules. For example, $d_\sigma d_\sigma = \sum_c N^c_{ab} d_c$ would give $d_\sigma = \sqrt{2}$.

We have the rules for setting up our Bose-Mesner algebra with Schur product and we refer to the elements of the algebra with the same $\sigma, \psi, 1$ notation. The fusion rules of equation \eqref {FusionRules} are the Krein parameters of the algebra that describe the adjacency 3x3 matrix of a family of four possible graphs $\mathscr{A} = \{A_1, A_2, A_3, A_4\}$ along with the unit element. For example, $A_1 = \begin{bmatrix} 1 & 0 & 0 \\ 0 & 0 & 0 \\ 0 & 0 & 0 \end {bmatrix}$ and $A_2 \begin{bmatrix} 0 & 1 & 0 \\ 1 & 0 & 0 \\ 0 & 0 & 0 \end {bmatrix}$ and of course these two are orthogonal under Schur multiplication. The vertices of the graphs are colored according to that of the anyon type and this information will be used to define the q-deformed states, $q_\sigma = e^{i\frac{\pi}{8}}, q_\psi = -1$. The polynomials $A^n_i, i = 1, ..., 4$ encode the n-distance paths of the quantum walks of the anyons that is equivalent to cascading pair-of-pants n-times. Our quantum probability space is $(\mathscr{A}, \rho)$ where $\rho$ is a linear map on the algebra $\mathscr{A}$ satisfying some regularity conditions. For example, a state can assign the subgraphs $N^1_{\sigma \sigma} = 1$, $N^{\psi}_{\sigma \sigma} = -11$, and zero for the rest of the basis of the algebra, by linearity the map can be extended to the whole algebra, corresponding to a qubit in quantum information processing. If desired, a Hilbert space can be derived from the von Neumann algebra via the GNS construction and the basis of the algebra form the projections of the Hilbert space. For this state, the matrix algebra can be described by 2x2 matrices and the Braid matrices act as automorphisms.

By applying Theorem 1 from our earlier work \cite {Radbalu2020} we can construct a family of quantum Markov chains on the basis set of the Bose-Mesner algebra with the Schur product *.

Suppose we have a  commutative association scheme $\{A_j\}_{j=0}^d$. Then, we can simultaneously diagonalize the matrices $A_0,\dotsc,A_d$ by the spectral theorem. Therefore, the adjacency algebra $\mathscr{A}$ has an alternative basis $E_0,\dotsc,E_d$ of projections onto the maximal common eigenspaces of $A_0,\dotsc,A_d$. Since $\mathscr{A}$ is closed under the Schur (Hadamard) product, there are coefficients $q_{i,j}^k$ such that
\[ E_i \circ E_j = \frac{1}{|X|} \sum_{k=0}^d q_{i,j}^k E_k \qquad (0 \leq i,j \leq d). \]
The coefficients $q_{i,j}^k$ are called the \emph{Krein parameters} (fusion rules in our case) of the association scheme. This leads to a commutative hypergroup. Let $m_j = \rank E_j$, and define $e_j = m_j^{-1} E_j$. Then
\[ e_i \circ e_j = \frac{1}{|X|}\sum_{k=0}^d \left( \frac{m_k}{m_i m_j} q_{i,j}^k \right) e_k. \]

The dual notion to Krein parameters is the Intersection numbers $p^k_{ij}$ in terms of the usual matrix product
$A_i \bullet A_j= \sum_{k} p^k_{ij}A_k$.
Intuitively, this means in a distance-regular graph (ex: complete graphs, cycles, and odd graphs) the number of paths between a pair of k-distant vertices via i-distant plus j-distant paths is independent of the pair.
\begin{theorem}
For each $i,j$, the mapping $k\mapsto \frac{m_k}{m_i m_j} \frac{q_{i,j}^k}{|X|}$ defines a probability distribution $\mu$ on $\{0,\dotsc,d\}$. If we define
\[ (e_i * e_j)(k) = \frac{m_k}{m_i m_j} \frac{q_{i,j}^k}{|X|}, \]
so that
\[ e_i \circ e_j = \sum_{k=0}^d ( e_i * e_j)(k)\cdot e_k, \]
then $\{e_0,\dotsc,e_d\}$ has the structure of a commutative hypergroup with identity element $e_0 = \frac{1}{|X|}J$ (the all-ones matrix, scaled by $|X|^{-1}$) and involution given by entry-wise complex conjugation.
\end{theorem}
A probability measure is characterized by m-moments $\forall m \ge 1$ and in the context of graphs they correspond to m-step walks from starting and ending at the same vertex. The above result prescribes a classical Markov chain canonically induce and a quantum versions can be constructed as follows:

Fix any $e_i$, and define $T\colon \mathscr{A} \to \mathscr{A}$ by $T(M) = e_i \circ M$ (Hadamard multiplication) for any $M \in \mathscr{A}$. Since $e_i$ is positive, $T$ is completely positive \cite[Theorem~3.7]{Paulsen2002}. Moreover, $\mathscr{B}$ is an invariant subspace of $T$, and $\left. T\right|_{\mathscr{B}}$ describes a classical Markov chain on the state space $e_0,\dotsc,e_{d'}$, corresponding to a random walk on the hypergroup $\{e_0,\dotsc,e_{d'}\}$.
we can build a family of Quantum Markov Chains indexed by $\{0\leq{i}\leq{d}\}$. we could replace $e_i$ with any convex combination of $e_0,\dotsc,e_d$ and obtain another chain.
In terms of quantum walks, we recommend the work of Wang et al \cite {Wang2010} to get physical picture of how such an evolution can be fashioned, we can think of $e_d$ is the coin operator and we can apply the unitaries of the Braid group to it as rotations for simulating all the varied processes of the powerful framework.
As the chain moves around the state space the graph grows in size (sometimes decreases in size) by cascading pair-of-pans and once we identify the stochastic independence, in future work, the correct quantum central limit theorem may be applied. By fixing one of the elements of the hypergroup we mean choosing a state and then the evolution is viewed in Heisenberg picture. In the Ising system we can imagine the Z space is represented by the $\psi$ particles that are fixed in positions, with the advantage that countable anyons can encode states exponentially more efficiently than any other representation of Z, and $\sigma$ acts as the coin with the integer line generated dynamically as the walk evolves. The probability amplitudes accumulated on the fixed anyons can provide interesting statistical ensembles.
\end {example}

In our BM algebra the Schur multiplication or the Hadamard product, it consisting of multiplying the corresponding elements of the matrix, plays an important role. One way to understand the connection between Hadamard operation and entanglement is to consider the identity of the operation and view it as the projection onto a maximally entangled state in a chosen basis. More formally, the identity of the Schur multiplication is given by 
\begin {equation}
E = \sum_{i,j} e_{ij} = \sum_{i,j} \ket{e_j}\bra{e_i} = \ket{\sum_i e_i}\bra{\sum_j e_j} = d\ket{e}\bra{e} .
\end {equation}
In the above, $\bra{e} = \frac{1}{\sqrt{d}}\sum_j e_j$. More on this product to develop intuition on our constructions will be provided later.  

Next, let us consider the Fibonacci system of anyons and build entangled Markov chains using the second theorem in our previous work \cite {Radbalu2020}.
\begin {example} It consists on a single non abelian anyon $\mathscr{L} = 1, f$ with the fusion rule f x f = 1 + f and the system supports universal computing. An family of entangled Markov chains $\hat{\mathbb{E}}^i$ indexed by can be defined as
\begin {equation} \label {eq: TE}
\hat{\mathbb{E}}^i(M\otimes{N}) = m \circ [M\otimes{P(N)}], \{0\leq{i}\leq{d}\}
\end {equation}
where $\hat{\mathbb{E}}$ is the transition expectation, quantum analogue of classical transition operator, $\otimes$ is the Hadamard product, and P is the probability transition matrix of a classical chain. With this propagator we have chain that entangles the sites as it evolves while embedding the classical chain. To realize this chain in anyonic set up we have to first encode the classical probability transition matrix $P = ((p_{ij}))$ as an unitary operator. One way to construct such a unitary is as follows:
\begin{equation}
U=\left( \begin{array}{cccc}
             p_0^{1/2} & p_1^{1/2 }  & \dots &p_{d-1}^{1/2}\\
            -p_1^{1/2}  &{ }  &{ }  &{ }\\
             \vdots &1  &-  &Q\\
            -p_{d-1}^{1/2}  &{ } &{ } &{ }
             \end{array} \right) \text{, where   } Q=((q_{ij})),q_{ij}=\frac{(p_{i}p_j)^{1/2}}{(1+p_0^{1/2})}, i,j\geq{1}
\end{equation}
Now, we can use braiding to realize this unitary and apply fusion with $e_i$ to fashion the chain dynamics. 

To develop insight into our constructions let us consider the transition expectation $\hat{\mathbb{E}}^i$ is generated by the isometry \cite {Radbalu2020} 
\begin {align*}
V\ket{e_{ij}} &= \sum\limits_{j_{n+1}\in{S}}\sqrt{t_{ij}}\ket{e_i}\otimes\ket{e_j}. \\
V^*\ket{e}_i\bra{e}_j &= \sqrt{t_{ij}}\ket{e}_i.
\end {align*} 
as $\hat{\mathbb{E}} (X) = V^* X V$. It is easy to see this as same as the form in equation \eqref {eq: TE} that is in terms of Schur multiplication which is not only an elegant description but it has physical interpretation as fusion rules. 

The same operator we have described in our earlier work \cite {RadLiu2017} in matrix form A that plays an important role with its spectrum forming stationary states of the Markov chain when they exist. It can be used to construct quantum walk propagator that embeds a classical Markov chain \cite {Szegedy2004}.
\begin {align*} \nonumber
A^\dag A &= I. \\
AA^\dag &= \Pi.     \text {  Projection Operator} \\
S &= (A^\dag)^{-1} D A^{-1}. \text {  Swap Operator}  \\
U &= S(2\Pi - I).  
\end {align*}
Here, again we embedded the classical Markov chain, described by the transition probability matrix D, in the quantum counterpart. The eigen space of the matrix D when lifted to the quantum space via the matrix transformation A provides the invariant subspace under the unitary evolution provide by the Grover like diffusion operator. The operator A acts on a Hilbert space $\mathscr{H}_e$ of pairs of vertices of a graph whose vertices form the state space $\mathscr{H}_v$ of the classical Markov chain. We described two subspaces $\mathscr{H}_\psi, \mathscr{H}^\perp_\psi$ of $\mathscr{H}_e$ that are invariant with respect to the quantum walk unitary propagator. In our formalism in this work the Schur multiplication on adjacency matrices enforces the same invariance and restricts evolution on connected vertices. In addition, the fusion rules interpretation of the product makes it easier to implement physically using anyonic systems. To describe the quantum walk unitry U in terms of adjacency matrices we have to construct an interacting Fock space (IFC), as we have done here \cite{Radbalu2020}, that we will take up in a future study.
\end {example}

\section {Summary and Conclusions}
Starting with the fusion rules of a unitary modular category we constructed quantum Markov chains that evolve on the hypergroups of an association scheme. We then synthesized entangled versions of the QMCs with an embedded classical chain. We provided examples of anyonic systems described in association schemes and quantum probabilistic framework generating another perspective for topological quantum computation.
\
\section {Acknowledgements}
\section {Appendix}
\section {Association Schemes}
\begin{defn}
Let $X$ be a (finite) vertex set, and let $\mathfrak{X} = \{A_j\}_{j=0}^d$ be a collection of $X\times X$ matrices with entries in $\{0,1\}$. We say that $\mathfrak{X}$ is an \emph{association scheme} if the following hold:
\begin{enumerate}[(1)]
\item $A_0 = I$, the identity matrix;
\item $\sum_{j=0}^d A_j = J$, the all-ones matrix (In other words, the $1$'s in the $A_j$'s partition $X\times X$);
\item For each $j$, $A_j^T \in \mathfrak{X}$; and
\item For each $i,j$, $A_i A_j \in \spn\mathfrak{X}$.
\end{enumerate}
A \emph{commutative} association scheme also satisfies
\begin{enumerate}[(1)]
\setcounter{enumi}{4}
\item For each $i,j$, $A_i A_j = A_j A_i$.
\end{enumerate}
\end{defn}
The above association scheme may be viewed as the adjacency matrices of graphs with a common set of $|\mathfrak{X}| = d$ vertices. Alternately, the scheme can represent 1-distance, 2-distance, ..., d-distance matrices of the same graph. We will take the former view while discussing multi-modal interacting Fock spaces later.
\begin{example} \label{ex:1}
Let $X=G$ be a finite group. For each $x\in G$, let $A_x$ be the the matrix for left translation by $x$ in $\ell^2(G)$. In other words, $A_x$ is the $G\times G$ matrix with
\[ (A_x)_{y,z} = \begin{cases}
1, & \text{if }y=xz \\
0, & \text{otherwise}
\end{cases} \]
for $y,z \in G$. When $e\in G$ is the identity element, we have $A_e = I$, $\sum_{x\in G} A_x = J$, $A_x^T = A_{x^{-1}}$, and $A_x A_y = A_{xy}$. Thus, $\mathfrak{X}:= \{A_x\}_{x\in G}$ is an association scheme.
\end{example}

\begin{example}
\item Let $G$ be a finite group acting transitively on a finite set $X$. Then $G$ also acts on $X\times X$ through the action $g\cdot (x,y) = (g\cdot x, g\cdot y)$ for $g\in G$ and $x,y \in X$. Let $R_0,\dotsc,R_d \subset X\times X$ be the orbits for this action, numbered so that $R_0 = \{ (x,x) : x \in X\}$. (This is an orbit since $G$ acts transitively on $X$.) For each $j=0,\dotsc,d$, let $A_j$ be the $X \times X$ matrix with
\[ (A_j)_{x,y} = \begin{cases}
1, & \text{if } (x,y) \in R_j \\
0, & \text{otherwise.}
\end{cases} \]
Then, one can show, $\mathfrak{X} = \{A_j\}_{j=0}^d$ is an association scheme. It will be commutative if and only if the action of $G$ on $X$ is \emph{multiplicity free}. In other words, the permutation representation of $G$ associated with its action on $X$ decomposes as a direct su, of irreducibles, with no irreducible repeated up to unitary equivalence.
\end{example}

\begin{example} \label{ex:3}
Let $X=G$ be a finite group, and let $K \subset \Aut(G)$ be a group of automorphisms of $G$. Let $\{e\} = C_0,\dotsc, C_d$ be the orbits of $K$ acting on $G$. If $\{A_x\}_{x\in G}$ are as in Example~\ref{ex:1}, define $B_0,\dotsc,B_d$ by
\[ B_j := \sum_{x\in C_j} A_x. \]
Then $\mathfrak{X}:=\{B_j\}_{j=0}^d$ is an association scheme. We call this a \emph{subscheme} of $\{A_x\}_{x\in G}$.

When $K$ is the group of inner automorphisms of $G$ (i.e.\ conjugations by elements of $G$), the orbits $C_0,\dotsc,C_d$ are precisely the conjugacy classes of $G$. Then $\mathscr{B}:=\spn\{B_0,\dotsc,B_d\}$ is the center of the group von Neumann algebra $\mathscr{A}:=\spn\{A_x : x \in G\}$.
\end{example}

\begin {example} The Johnson scheme J (v,k). The vertex set of this scheme is the set of all k-subsets of a fixed set of v elements. Two vertices $\alpha$ and $\beta$ are i-related if $\|\alpha \cap \beta\| = k - i$. This scheme has k classes.
\end {example}

\begin {example} \label {ex: Grassmann}
The Grassmann scheme $J_q (v, d)$. The vertex set is the set of all subspaces of dimension d of the vector space of dimension n over GF(q) (finite field with q elements). Subspaces $\alpha$ and $\beta$ are i-related if $dim(\alpha \cap \beta) = i$. This q-deformed Johnson scheme has d classes, may be thought of as a discrete version of a Grassmannian manifold, and the graph it generates is distance transitive and the basis for our construction of an IFS.
\end {example}
\begin{defn}
The \emph{adjacency algebra} of an association scheme $\{A_j\}_{j=0}^d$ is $\mathscr{A}:=\spn\{A_j\}_{j=0}^d$. Sometimes this is also called the \emph{Bose-Mesner algebra}. It's a unital $*$-algebra of matrices, i.e.\ a von Neumann algebra. It is also closed under the Hadamard (Schur) product.
\end{defn}

\bibliographystyle{abbrv}

\end{document}